\documentclass[aps, 10pt, prd, notitlepage, twocolumn, superscriptaddress,nofootinbib, floatfix]{revtex4-1}
\usepackage[pdftex]{graphicx}              
\usepackage{amsmath}
\usepackage{amssymb}
\usepackage{amsfonts}
\usepackage[utf8]{inputenc}
\usepackage[T1]{fontenc}
\usepackage{mathrsfs}
\usepackage{anyfontsize}

\usepackage[linktocpage,breaklinks]{hyperref}
\usepackage[usenames,dvipsnames]{xcolor}

\usepackage{tensor}
\usepackage{bm}

\usepackage{graphicx}
\usepackage{epsfig}
\usepackage{epstopdf}

\usepackage{natbib}
\usepackage{hyperref}
\usepackage{verbatim}

\hypersetup{colorlinks=true,
            citecolor=RoyalBlue,
            linkcolor=RoyalBlue,
            urlcolor=RoyalBlue}

\newcommand{\be}{\begin{eqnarray}}
\newcommand{\ee}{\end{eqnarray}}

\begin{document}

\title{Modeling uncertainties in X-ray reflection spectroscopy measurements II:\\Impact of the radiation from the plunging region}

\author{Alejandro C\'ardenas-Avenda\~no} 
\affiliation{Programa de Matem\'atica, Fundaci\'on Universitaria Konrad Lorenz, 110231 Bogot\'a, Colombia}
\affiliation{Department of Physics, University of Illinois at Urbana-Champaign, Urbana, Illinois 61801, USA}

\author{Menglei~Zhou}
\affiliation{Center for Field Theory and Particle Physics and Department of Physics, Fudan University, 200438 Shanghai, China}

\author{Cosimo~Bambi}
\email[Corresponding author: ]{bambi@fudan.edu.cn}
\affiliation{Center for Field Theory and Particle Physics and Department of Physics, Fudan University, 200438 Shanghai, China}

\date{\today}


\begin{abstract}
X-ray reflection spectroscopy is a powerful tool to probe the strong gravity region around black holes, but the available relativistic reflection models have a number of simplifications that lead to systematic uncertainties (not fully under control) in the measurement of the properties of a source. In Paper~I, we considered the case of an optically thin plunging region and we studied the impact of the radiation produced by the other side of the disk or circling the black hole one or more times. In the present paper, we discuss the case of an optically thick plunging region and we study the impact of the reflection spectrum of the plunging gas. We show that the contribution of such radiation is more important for low and negative values of the black hole spin parameter and large values of the viewing angle, and it decreases significantly as the spin parameter increases and the inclination angle decreases. While the estimate of some parameters may be affected by the reflection spectrum of the plunging gas if this is not included in the theoretical model, we find that such radiation does not appreciably limit our capability of testing the Kerr black hole hypothesis.  
\end{abstract}


\maketitle

\section{Introduction}

Relativistic reflection features are commonly observed in the X-ray spectra of black hole binaries and active galactic nuclei (AGNs)~\cite{Tanaka:1995en,Nandra:2007rp,Miller:2007tj,Walton:2012aw,Risaliti:2013cga,Miller:2013rca,Tomsick:2013nua,Bambi:2017iyh}. They are produced when the accretion disk is illuminated by a hot corona~\cite{George:1991jj,Ross:2005dm,Garcia:2010iz}. X-ray reflection spectroscopy is the analysis of the relativistic reflection spectrum of an accreting black hole and can be a powerful tool to study the inner part of the accretion disk~\cite{DeRosa:2018aka}, measure black hole spins~\cite{Brenneman:2006hw,Miller:2009cw,Reynolds:2013qqa,Miller:2014aaa}, and even test Einstein's theory of general relativity (GR) in the strong field regime~\cite{Schee:2008fc,Johannsen:2012ng,Bambi:2012at,Bambi:2015kza,Bambi:2016sac,Cao:2017kdq,Tripathi:2018lhx, Tripathi:2019bya, Zhang:2019ldz, Cardenas-Avendano:2019zxd}.

As in any astrophysical measurement, accurate and precise estimates of the properties of a source with X-ray reflection spectroscopy require sufficiently sophisticated theoretical models to limit modeling bias. The available relativistic reflection models have a number of simplifications that introduce systematic uncertainties in the final measurements~\cite{Liu:2019vqh,Riaz:2019bkv,Riaz:2019kat, Cardenas-Avendano:2019pec}. In order to get more reliable measurements of accreting black holes using X-ray reflection spectroscopy, it is thus crucial to develop more sophisticated theoretical models, removing the unjustified model simplifications, and, at the same time, have at least a rough estimate of the impact of all the model simplifications on the measurement of the parameters of interest to ensure that the associated systematic errors are much smaller than the statistical errors of the measurement.

Simplifications in the relativistic reflection models can be conveniently grouped into four classes: $i)$ simplifications in the calculation of the reflection spectrum at the emission point in the rest-frame of the gas, $ii)$ simplifications in the description of the accretion process, $iii)$ simplifications in the description of the hot corona, and $iv)$ simplifications related to relativistic effects not taken into account. Depending on the intrinsical physical features of the source during the observation, the properties of the observational facility, and the aim of the study, some simplifications are more or less justified than others.

In Ref.~\cite{Zhou:2019dfw} (Paper~I), accreting black holes with optically thin plunging regions, namely the region between the black hole event horizon and the inner edge of the accretion disk, were considered. In such a situation, the distant observer also sees the reflection radiation produced by the other side of the disk as well the reflection radiation circling the black hole one or more times (higher order disk images). Then observations were simulated with the XIFU instrument of \textsl{Athena}~\cite{Nandra:2013jka} and fitted with a model that did not include the calculation of the reflection spectrum produced by higher order disk images in order to study the capability of the model to recover the correct input parameters, with particular attention to the possibility of testing the Kerr metric around the source. In Paper~I, it was found that the effect of higher order disk images can be safely ignored for observations with present and near future X-ray facilities.

In the present paper (Paper~II), we consider the opposite case, namely accreting black holes with optically thick plunging regions, and we study the impact of the reflection radiation of the plunging gas on X-ray reflection spectroscopy measurements, with particular interest on tests of the Kerr black hole hypothesis and the measurement of the deformation parameters of the background metric. For a steady-state, axisymmetric, geometrically thin disk, we can employ the continuity equation to derive the optical depth to electron scattering and we find (using units in which $G_{\rm N} = c = 1$)~\cite{Reynolds:1997ek}
\be\label{eq-tau}
\tau_{\rm e} = \frac{2}{\eta | u^r |} \left( \frac{r_{\rm g}}{r} \right) 
\left( \frac{L}{L_{\rm Edd}} \right)
\ee
where $L = \eta \dot{M}$ is the accretion luminosity, $\eta$ is the radiative efficiency, $L_{\rm Edd}$ is the Eddington luminosity, $r_{\rm g} = M$ is the gravitational radius, and $u^r$ is the radial component of the 4-velocity of the accreting gas in the plunging region. From Eq.~(\ref{eq-tau}), we see that the plunging region is optically thin for very low mass accretion rates and optically thick otherwise: $\tau < 1$ for $L/L_{\rm Edd} < 0.05$ in the Schwarzschild spacetime with $\eta = 0.06$, and for $L/L_{\rm Edd} < 0.01$ in the Kerr spacetime with $a_* = 0.998$ and $\eta = 0.3$.

As a preliminary study to estimate the magnitude of the impact of the reflection radiation produced in the plunging region, and because of our ignorance on the geometry of the corona and the properties of the gas in the plunging region, we employ some toy models to calculate reflection spectra taking into account the contribution from the plunging gas. In order to evaluate the impact of the radiation from the plunging region on current and near-future X-ray reflection spectroscopy measurements, and in particular on the tests of the Kerr metric, we proceed as in Paper~I, and we simulate some observations with the XIFU instrument of \textsl{Athena}. We then fit the data with the model {\sc relxill\_nk}~\cite{Bambi:2016sac,Abdikamalov:2019yrr} and we compare the best-fit values with the known input parameters.  We find that the contribution of such radiation is more important for low and negative values of the black hole spin parameter and large values of the viewing angle, and it decreases significantly as the spin parameter increases and the inclination angle decreases.

This paper is organized as follows:
Sect.~\ref{ISCO} summarizes the accretion disk model used and the general expressions that govern the fluid in the plunging region;
Sect.~\ref{Spectra} presents the modifications on single iron lines and on full reflection spectra of the considered model;
Sect.~\ref{Simulations} shows the results of the different type of simulations performed;
Sect.~\ref{Conclusions} is for the discussion of our results and some concluding remarks. 
Throughout the paper, we mostly use geometric units in which $G_{\rm N}=c=1$, and the $(-,+,+,+)$ metric signature. 
Commas in index lists will stand for partial derivatives.

\section{Radiation from the plunging region}
\label{ISCO}

Our systems are black holes accreting from geometrically thin and optically thick disk, and we employ the Novikov–Thorne model~\cite{novikov1973astrophysics,Page:1974he}. The disk lies on the equatorial plane of the system, perpendicular to the black hole spin axis. The particles of the gas in the disk move on nearly-geodesic, equatorial, circular orbits (Keplerian motion). As the gas loses energy and angular momentum, it slowly inspirals towards the black hole. When the gas reaches the inner edge of the disk, here assumed at the innermost stable circular orbit (ISCO)~\cite{Bardeen:1972fi}, it plunges onto the black hole. The plunging region is the region between the inner edge of the disk and the black hole event horizon. Depending on its gas density, the plunging region can either be optically thin (very low mass accretion rate) or optically thick (otherwise). If the plunging region is optically thin, because of the strong light bending in the vicinity of the black hole, the observed reflection spectrum far from the source receives the contributions from the reflection spectrum produced by the other side of the disk or circling the black hole one or more time. This was the scenario studied in Paper~I. On the other hand, if the plunging region is optically thick, we should observe the reflection spectrum produced by illumination of the plunging gas by the hot corona. This is the scenario investigated in the rest of this paper.

For the description of the flux detected by an observer, we follow Ref.~\cite{Bambi:2016sac}, which applies the formalism of the transfer function for geometrically thin and optically thick accretion disks around a black hole~\cite{Cunningham:1975zz}. The transfer function in this context can be interpreted as an integration kernel to calculate the spectrum detected by the distant observer, starting from the local spectrum at any point of the disk. The usefulness of this approach is that the observed flux can be rewritten as a function that is proportional to the transfer function $f (g^*, r_{\rm e}, i)$, which depends on the spacetime metric, the disk model, and the viewing angle of the distant observer, and the specific intensity of the radiation as measured by the emitter $I_{\rm e}(\nu_{\rm e},r_{\rm e},\vartheta_{\rm e})$, which depends on atomic physics and the disk intensity profile. Therefore, the observed flux is given by~\cite{Bambi:2016sac}
\be
F_{\rm o} (\nu_{\rm o}) 
&=& \frac{1}{D^2} \int_{r_{\rm in}}^{r_{\rm out}} \int_0^1
 \frac{\pi r_{\rm e} g^2 f(g^*,r_{\rm e},i)
I_{\rm e} (\nu_{\rm e},r_{\rm e},\vartheta_{\rm e})}{\sqrt{g^* (1 - g^*)}}
 \, dg^* \, dr_{\rm e}  \, , \nonumber\\
\label{eq-Fobs}
\ee
where $D$ is the distance of the observer from the source, $r_{\rm in}$ and $r_{\rm out}$ are, respectively, the inner and the outer edges of the disk, $r_{\rm e}$ is the emission radius in the disk, $\vartheta_{\rm e}$ is the emission angle (i.e., the angle between the normal to the disk and the photon propagation direction), $i$ is the viewing angle (i.e., the angle between the normal to the disk and the line of sight of the observer), $g = \nu_{\rm o}/\nu_{\rm e}$ is the redshift factor, $\nu_{\rm o}$ is the photon frequency as measured by the distant observer, and $\nu_{\rm e}$ is the photon frequency in the rest frame of the emitter. The relative redshift factor $g^*$ (which ranges from 0 to 1) is defined as $g^* = \left( g - g_{\rm min} \right) / \left(g_{\rm max} - g_{\rm min} \right)$, where $g_{\rm max}=g_{\rm max}(r_{\rm e},i)$ and $g_{\rm min}=g_{\rm min}(r_{\rm e},i)$ are, respectively, the maximum and the minimum values of the redshift factor $g$ for the photons emitted from the radial coordinate $r_{\rm e}$ and detected by a distant observer with viewing angle $i$. The redshift factor explicitly reads
\be\label{eq-ggeneric}
g = \frac{-u^\mu_{\rm o} k_\mu}{-u^\nu_{\rm e} k_\nu} \, ,
\ee
where $u^\mu_{\rm o} = (1,0,0,0)$ is the 4-velocity of the distant observer, $k^\mu$ is the 4-momentum of the photon, and $u^\nu_{\rm e} $ is the 4-velocity of the particles of the gas. The material in the disk follows nearly-geodesic, equatorial, circular orbits, slowly spiraling inwards until it reaches the ISCO at $r = r_*$. We can write $u^\nu_{\rm e} = u^t_{\rm e} (1,0,0,\Omega)$, where $\Omega$ is the angular velocity $\Omega = d \phi/ d t$ measured by the distant observer. Assuming the Kerr background, we get 
\be\label{eq-g}
g^{\left(r \geq r_{*}\right)}=\frac{\sqrt{1-\left(r^{2}+a^{2}\right)\Omega^{2}-\frac{2M}{r}\left(1-a\Omega\right)^{2}}}{1+\lambda\Omega} \, ,
\ee
where $M$ and $a$ are, respectively, the mass and the rotational parameter of the black hole (the dimensionless spin parameter is $a_* = a/M$), and $\lambda \equiv k_{\phi}/ k_{t}$, is a constant of motion along the photon trajectory and can be evaluated from the initial conditions~\cite{Bambi:2012tg}. 

At $r= r_{*}$, circular geodesics become unstable and the material falls into the black hole with the energy $E_{*}$ and angular momentum $L_{*}$ of the geodesic of marginal stability~\cite{Speith_1995}. The 4-velocity of the gas in the plunging region is 
\be\label{eq-4velISCO}
u_{e}^{\nu}=\left(u_{e}^{t},u_{e}^{r},0,u_{e}^{\phi}\right) \, ,
\ee
where 
\begin{widetext}
\be\label{eq-4velISCOExplicit}
u_{e}^{t}	&=&	\frac{a^{3}\sqrt{M}r+a^{2}\sqrt{r_{*}}(-2Mr+2Mr_{*}+rr_{*})+a\sqrt{M}\left(r^{3}-2Mr_{*}^{2}\right)+r^{3}\sqrt{r_{*}}r_{*}-2M)}{r\left(a^{2}+r(r-2M)\right)\sqrt{2a\sqrt{Mr_{*}^{3}}-3Mr_{*}^{2}+r_{*}^{3}}} \\
u_{e}^{r}	&=&	-\frac{\sqrt{a^{2}+(r-2)r}\sqrt{-\frac{r(r-r_{*})^{2}\left(a^{2}(-(r+2r_{*}))+4a\sqrt{r_{*}}(r+r_{*})+r_{*}(r(r_{*}-4)-2r_{*})\right)}{\sqrt{r_{*}^{3}}\left(a^{2}+(r-2)r\right)\left(2a+(r_{*}-3)\sqrt{r_{*}}\right)}}}{r^{2}} \\
u_{e}^{\phi}	&=&	\frac{a^{2}r+2a\left(\sqrt{Mr_{*}^{3}}-r\sqrt{Mr_{*}}\right)+r_{*}^{2}(r-2M)}{r\left(a^{2}+r(r-2M)\right)\sqrt{\frac{2a\sqrt{Mr_{*}^{3}}-3Mr_{*}^{2}+r_{*}^{3}}{M}}}
\ee
\end{widetext}
and we get 
\be \label{eq-gISCO}
g^{\left(r < r_{*}\right)} =\frac{1}{u_{e}^{t}+\lambda u_{e}^{\phi}+\frac{k_{r}}{k_{t}}u_{e}^{r}}
\ee
where $k_{r}/k_{t} = - k^{r}g_{rr}/ k_{0}^{t}$ and $g_{rr}$ is the $\left(r,r\right)$ component of the metric.

The intensity profile of the accretion disk could be calculated theoretically if the coronal geometry were known. In the case of a corona with arbitrary geometry, it is common to employ an intensity profile described by a power law or a broken power law. In the simplest case of a power law, the emissivity in the disk ($r \geq r_{*}$) can be written as
\be \label{eq-P1}
\epsilon^{\left(1\right)} \left( r, q \right) \propto \left(\frac{r_{*}}{r}\right)^q.
\ee
where $q$ is the emissivity index. In the case of a point-like corona source along the black hole spin axis, the intensity profile of the disk reduces to the form in Eq.~(\ref{eq-P1}) with $q=3$ in the Newtonian limit (no light bending) at large radii.

\begin{figure}[htb]
\includegraphics[width=0.98\linewidth{}]{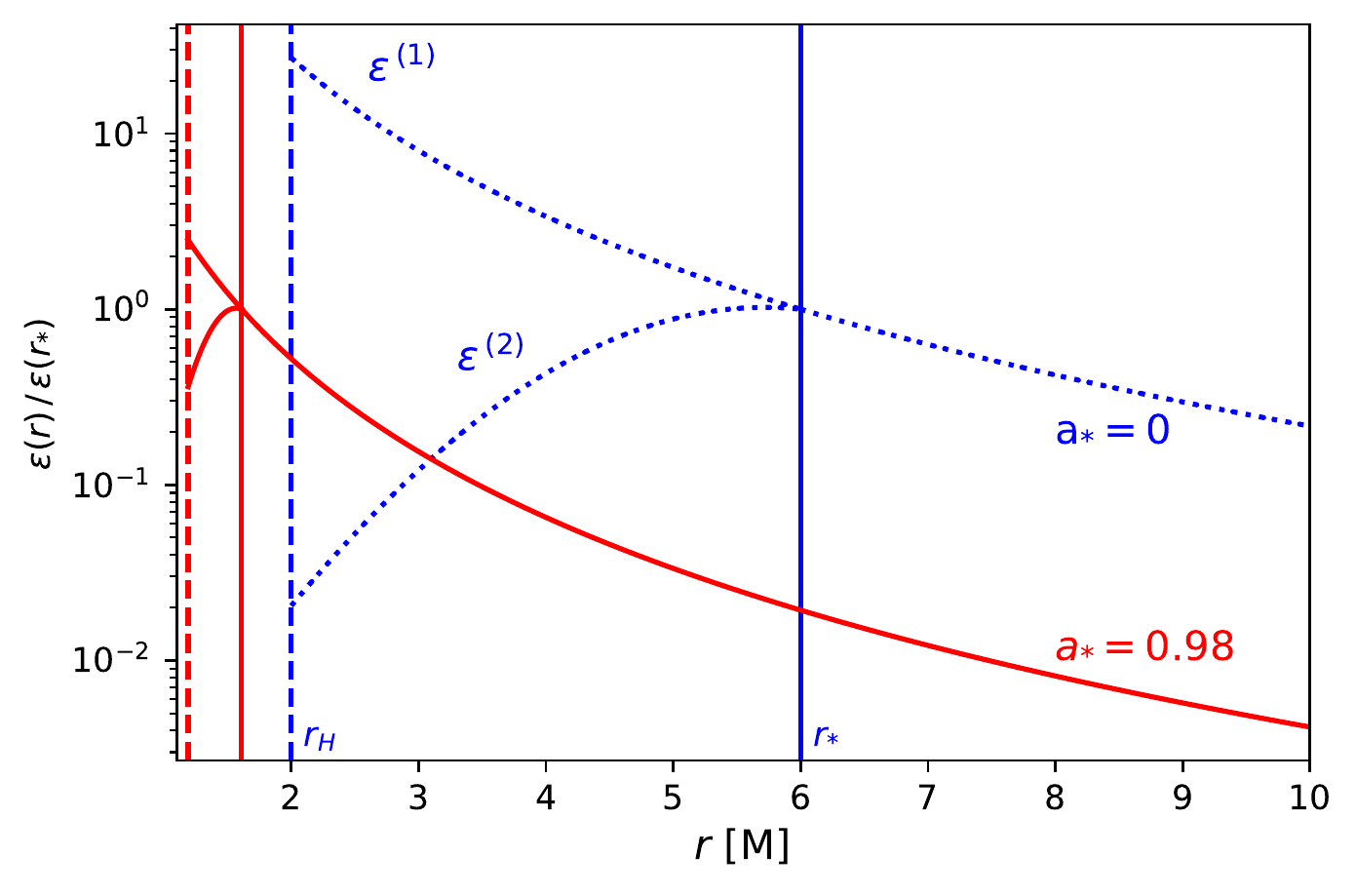}
\vspace{-0.4cm}
\caption{Emissivity profile for a Schwarzschild black hole (blue dotted lines) and for a Kerr black hole with spin parameter $a_*=0.98$ (red solid lines). In both cases, for $r \geq r_{*}$ we assume an emissivity profile described by a power law with $q=3$. For the plunging region $r< r_{*}$, we assume either the profile $\epsilon^{\,(1)}$ in Eq.~(\ref{eq-P1}) with $q=3$ or the profile $\epsilon^{\,(2)}$ in Eq.~(\ref{eq-P2}). The vertical solid lines denotes the location of the ISCO radius ($r_{*}=6\, M$ and $1.614\, M$, respectively). The dashed lines are drawn at the location of the horizon ($r_{H}=2\, M$ and $1.199\, M$, respectively). \label{f-profiles}}
\end{figure}

Concerning the intensity profile of the plunging region, $r< r{*}$, there is no common choice in literature since the radiation from the plunging region is normally ignored. In what follows, we consider two different emissivity profiles for the plunging region. The profiles we consider serve as the physical limits of the illumination profile of the accretion flow and should be taken merely as an approximation. Our profile~(1) assumes that the emissivity profile in Eq.~(\ref{eq-P1}) extends to the region inside the ISCO, leading to an appreciable illumination of the plunging region. Our profile~(2) is instead conservative, with a smooth transition that peaks at a radius inside the ISCO and then decays, and has the following form 
\be \label{eq-P2}
\epsilon^{\left(2\right)}  \left( r \right) \propto \left(\frac{r}{r_{*}}\right)e^{\frac{-\left(r - r_{*} \right)^{2}}{r_{*}-\ln{r_{*}-1}}}.
\ee
Fig.~\ref{f-profiles} sketches these profiles for a Schwarzschild black hole ($a_*=0$) and for a Kerr black hole with spin parameter $a_*=0.98$.

\section{Modification of the spectra}
\label{Spectra}

We now use the relativistic ray-tracing code presented in Refs.~\cite{Bambi:2016sac,Zhou:2019dfw} as a convolution model to transform a spectrum at the emission point of the disk to the observed spectrum far from the source, taking into account the relativistic effects. In the same spirit as the work done in Paper I, we perform an analysis of a single iron K$\alpha$ line and then of the full reflection spectrum. For the plunging region, we will study the two illumination profiles discussed in the previous section.  

 \subsection{K$\alpha$ line shapes}
 
 The most prominent feature in the reflection spectra of accreting black holes is often a broadened iron K$\alpha$ line. In the rest-frame of the gas, this is quite a narrow line at $6.4$~keV in the case of neutral iron and shifts up to $6.97$~keV for H-like iron (while there is no line for fully ionized iron). The iron K$\alpha$ line observed in the spectra of black holes is broadened and skewed as the result of the sum of all the contributions emitted from different parts of the accretion disk and differently affected Doppler boosting and gravitational redshift. As a first, very crude, approximation, we can model the reflection spectrum of the disk as a single iron K$\alpha$ line. The study of a single line can illustrate better the impact of the radiation from the plunging region, as the full spectrum is too convoluted, and makes it harder to clearly account for the qualitative impact.

\begin{figure*}[t]
\begin{center}
\includegraphics[]{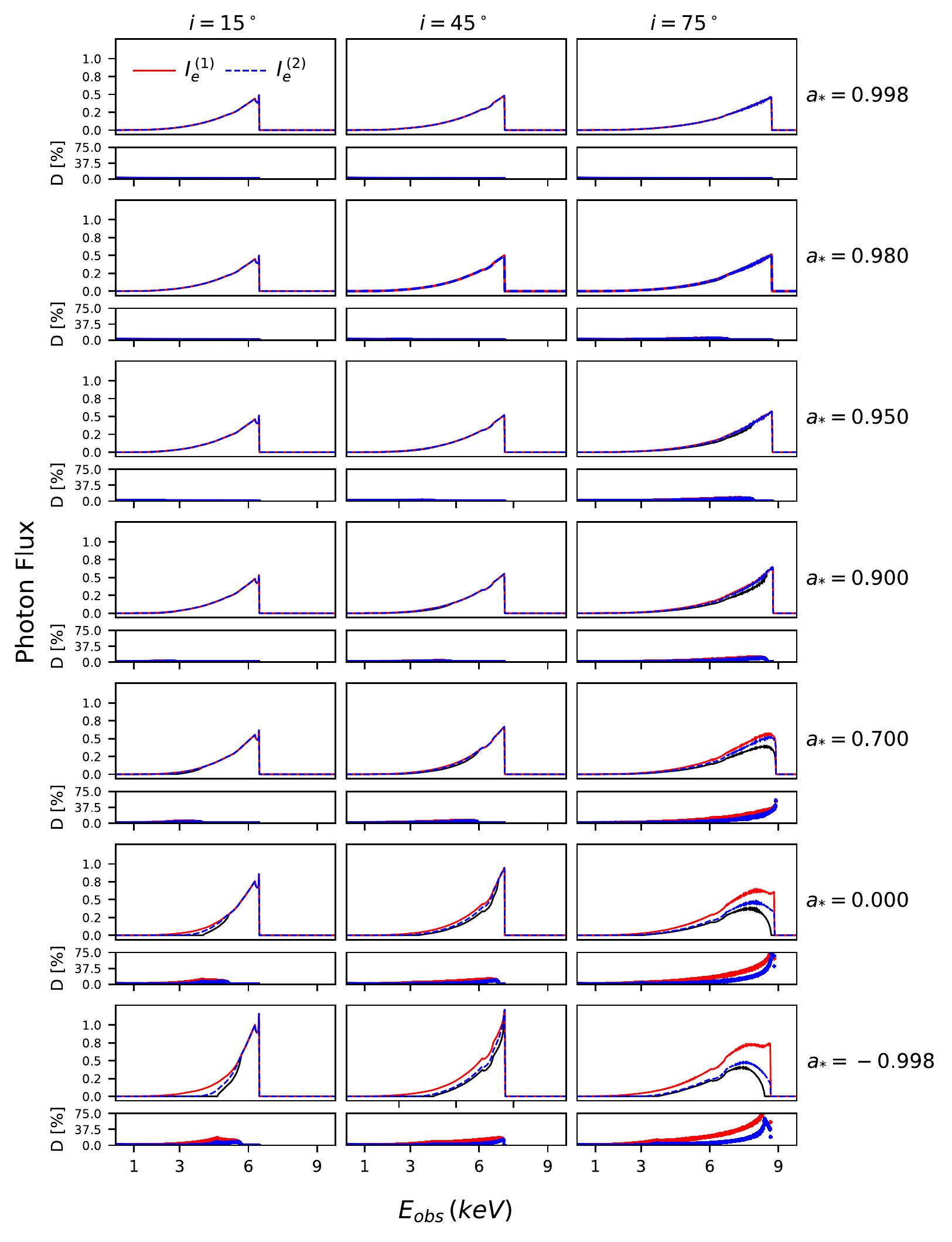}
\end{center}
\vspace{-0.8cm}
\caption{Iron line shapes for different black hole spins and viewing angles. In every panel, the top quadrant shows iron lines with (red and blue lines) and without (black lines) radiation from the plunging region. The red solid lines employ profile~(1) with $q=3$ for the plunging region and the blue dashed lines assume profile~(2). The intensity profile of the disk is always modeled by a power law with emissivity index $q=3$. The lower quadrants show the relative difference between the iron lines with and without radiation from the plunging region. \label{f-lines}}
\end{figure*}

We assume that the disk emission is a monochromatic line at a rest-frame energy $E_{0}=6.4$~keV and isotropic. The accretion disk region is $r_{*} \leq r \leq 400\rm{M}$ and has the emissivity profile $\epsilon^{\,(1)}$ in Eq.~(\ref{eq-P1}) with $q=3$. For the plunging region, $r_{\rm{H}} \leq r \leq r_{*}$, we consider profile~(1) in Eq.~(\ref{eq-P1}) with $q=3$ and profile~(2) in Eq.~(\ref{eq-P2}). The results are shown in Fig.~\ref{f-lines}, for different black hole spins and viewing angles.

The contribution of the radiation from the plunging region over the whole broadened line is clearly very small for high values of the spin parameter (with small plunging regions) and increases as the spin parameter decreases. The contribution from the plunging region is particularly important in the case of counterrotating disks (bottom panels in Fig.~\ref{f-lines}). The impact of the value of the viewing angle is similar, even if weaker: in this case, the contribution of the plunging region is more important for high values of the viewing angle and decreases as the viewing angle decreases. The relative difference between the lines with and without radiation from the plunging region is larger if we employ profile~(1) and smaller for profile~(2), but the difference is only at quantitative level, and the trend with black hole spins and viewing angles does not seem to depend on the choice of the emissivity profile in the plunging region.

 \begin{figure*}[t]
\begin{center}
\includegraphics[]{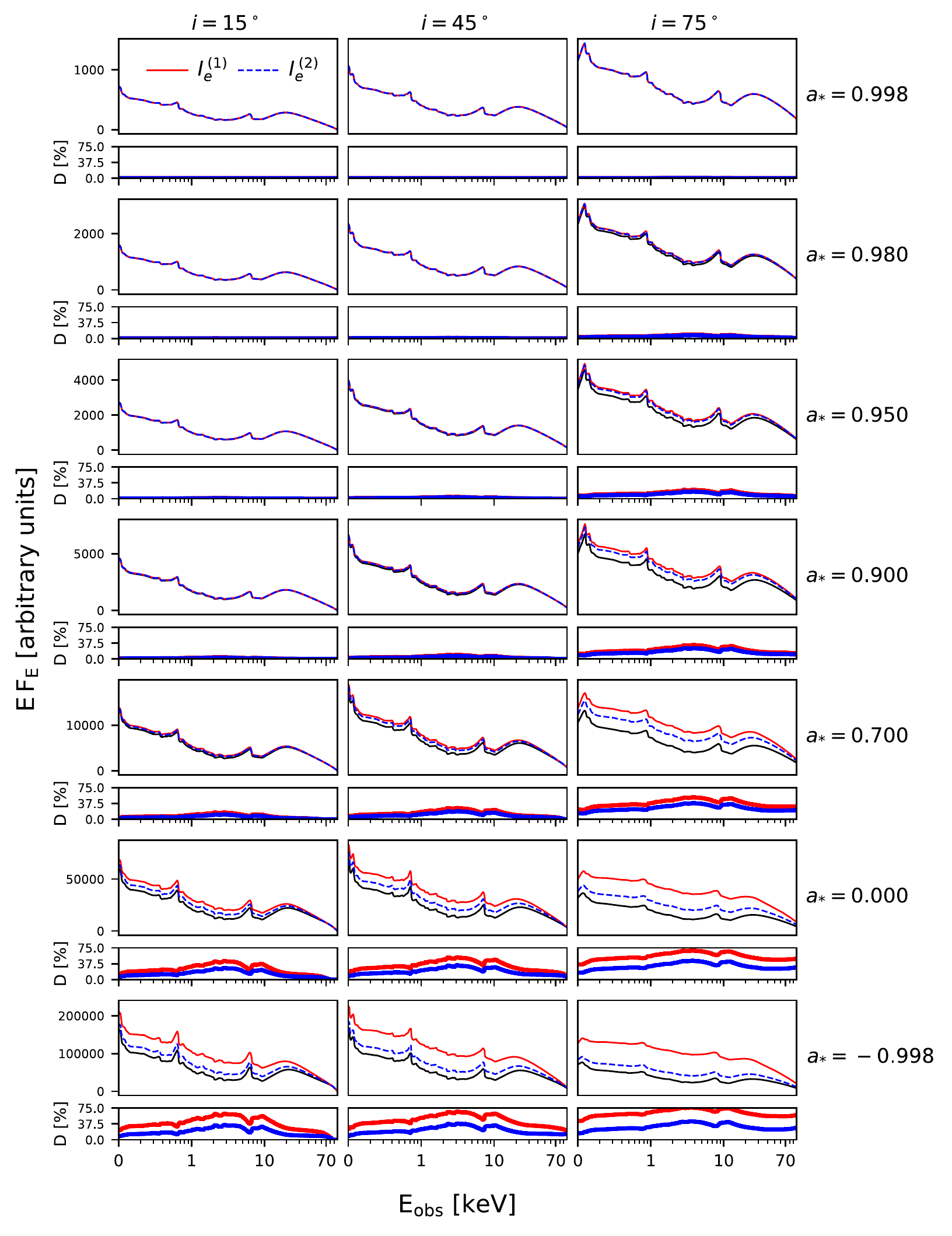}
\end{center}
\vspace{-0.8cm}
\caption{Reflection spectra for different black hole spins and viewing angles. In every panel, the top quadrant shows spectra with (red and blue lines) and without (black lines) radiation from the plunging region. The red solid lines employ profile~(1) with $q=3$ for the plunging region and the blue dashed lines assume profile~(2). In both cases, we employ the ionization parameter $\log\xi = 4.7$ for the material in the plunging region. The intensity profile of the disk is always modeled by a power law with emissivity index $q=3$. The lower quadrants show the relative difference between the spectra with and without radiation from the plunging region. \label{f-full}}
\end{figure*}

 \subsection{Full spectra}
 
We will now consider the whole reflection spectrum, which is extremely informative, given its many features, which are the heritage of the richness of the many physical processes occurring. Theoretical models to calculate the reflection spectrum in the rest-frame of the gas depend on the properties of the incident radiation from the corona (often modeled by a power law with an exponential cut-off, and thus specified by two parameters: the photon index $\Gamma$ and the high energy cut-off $E_{\rm cut}$), the properties of the accretion disk (usually specified by the iron abundance $A_{\rm Fe}$, measured in units of Solar iron abundance, and the ionization parameter $\xi$ or $\log\xi$, in units of erg~cm~$\rm{s}^{-1}$), and the emission angle $\vartheta_{\rm e}$. The ionization parameter $\xi$ is defined as
\be
\xi \equiv \frac{4 \pi F(r)}{n_{\rm e}(r)} \, ,
\ee
where $F ( r )$ is the X-ray flux illuminating a unit area of the disk at the radius $r$ and $n_{\rm e} ( r )$ is the comoving electron number density~\cite{Garcia:2013oma}.

For the non-relativistic reflection spectrum in the disk we use {\sc xillver}~\cite{Garcia:2013oma}, which provides a library of synthetic spectra for modeling the component of emission that is reflected from an illuminated accretion disk. We assume that the incident radiation has photon index $\Gamma=2$ and high energy cutoff $E_{{\rm cut}}=150$~keV. The accretion disk has Solar iron abundance, $A_{{\rm Fe}}=1$, and ionization parameter $\log\xi=3$. In the plunging region, the iron abundance must be the same, but a much higher ionization parameter is expected, which affects line emissions and continuum photoelectric absorption, mainly due to a rapid decrease of the electron density of the plunging region~\cite{Shafee:2008mm,Noble:2008tm,Wilkins:2020pgu}. For large values of $\xi$, fewer emission lines are present~\cite{Garcia:2013oma}.

Assuming the two different illumination profiles of the plunging region, (1) and (2), and a constant ionization parameter in the plunging region $\log\xi=4.7$ (the maximum allowed in {\sc xillver}), we found the spectra shown in Fig.~\ref{f-full} for different black hole spins and viewing angles. Since all the calculations in {\sc xillver} are done assuming the electron density $n_{\rm e}=10^{15}$~$cm^{-3}$, higher values of $\xi$ in the tables of {\sc xillver} correspond to higher X-ray fluxes $F ( r )$ increased by the same factor. Thus, in our simulations we need to renormalize the flux, such that they match at the ISCO and we are not getting an artificial higher flux in the plunging region. Fig.~\ref{f-full} shows the same trend already found in Fig.~\ref{f-lines} for the single line shapes. For rapidly rotating black holes and corotating disks, the contribution from the plunging region tends to be negligible, which can be easily interpreted with the fact that the size of the plunging region becomes smaller. However, some features show up, due to the change in the ionization parameter between the disk and the plunging region. In addition to an increase of the flux intensity, the spectrum tends to become smeared out and the contribution from the plunging region would be similar to a simple power-law continuum.

\section{Simulations}
\label{Simulations}

\begin{table}
\centering
{\renewcommand{\arraystretch}{1.3}
\begin{tabular}{lcccc}
\hline 
\hline
Simulation  \hspace{0.2cm} & $a_*$ & $i$ & Profile $(r < r_*)$  \hspace{0.2cm} & $\log\xi$ $(r < r_*)$ \\
\hline 
\hline
A1 & 0.9 & \hspace{0.2cm} $45^\circ$ \hspace{0.2cm} & (1) & 4.7 \\
A2 & 0.9 & $45^\circ$ & (2) & 4.7 \\
\hline
B1 & 0.9 & $75^\circ$ & (1) & 4.7 \\
B2 & 0.9 & $75^\circ$ & (2) & 4.7 \\
\hline
C1 & 0.99 & $75^\circ$ & (1) & 4.7 \\
C2 & 0.99 & $75^\circ$ & (2) & 4.7 \\
\hline
D1 & 0.99 & $75^\circ$ & (1) & 3.0 \\
D2 & 0.99 & $75^\circ$ & (2) & 3.0 \\
\hline 
\hline
\end{tabular}
}
\caption{Summary of the input parameters of the simulations in our study. \label{t-ss}}
\end{table}

In order to quantify the impact of the contribution of the material inside the ISCO, we will now perform different experimental relativity studies with accretion disk spectral observations. Following the classification presented in Ref.~\cite{Cardenas-Avendano:2019pec}, we will perform two types of studies, where the synthetic injections consider the contribution of the material inside the plunging region and were constructed within GR. This synthetic injection would then be to recover with two models, one built within GR and the other one outside GR, both without taking into account the radiation inside the ISCO. The idea of these two types of experiments is to consider the implications to parameter estimation when the radiation from the plunging region is neglected, which can lead to a systematic bias. 

We simulate observations using the \verb5fakeit5 command of XSPEC~\cite{1996ASPC..101...17A}, the response file and the background spectrum of the XIFU instrument of \textsl{Athena}~\cite{Nandra:2013jka}, and the theoretical spectra produced in the previous section. In XSPEC language, the total model is 

\vspace{0.2cm}

{\sc tbabs$\times$(powerlaw+ reflection)} , 

\vspace{0.2cm}

\noindent where {\sc tbabs} takes the galactic absorption into account~\cite{Wilms:2000ez}, {\sc powerlaw} describes a power law component with a high energy exponential cut-off and represents the direct spectrum from the corona, and {\sc reflection} denotes our theoretical reflection spectrum that includes the radiation from the plunging gas. As done in Paper~I, we assume an observation of a very bright AGN (or a moderately bright black hole binary) and we set the photon flux to $1.4 \times 10 ^{-10}$~erg~cm$^{-2}$~s$^{-1}$. We simulate an observation with an exposure time of 100~ks, which leads to a photon count of about 35.5~millions photons in the 1-10~keV energy band. Before the spectral analysis, the simulated spectra are rebinned to have a minimum of 30~photon counts per bin in order to apply the $\chi^2$-statistics.

We simulate 8~observations, whose name and main properties are summarized in Tab.~\ref{t-ss}. In all these simulations we assume a Kerr spacetime. The spin parameter is $a_* = 0.9$ (A1, A2, B1, B2) and 0.99 (C1, C2, C3, C4). The emissivity profile in the disk is always modeled with a power low with emissivity index $q=3$. For the plunging region, we assume either profile~(1) in Eq.~(\ref{eq-P1}) with $q=3$ (simulations with the number 1 in the name) or profile~(2) in Eq.~(\ref{eq-P2}) (simulations with the number 2 in the name). The viewing angle is $i = 45^\circ$ in simulations A1 and A2 and $i = 75^\circ$ in the other simulations. The ionization parameter in the disk is always $\log\xi = 3$, while than in the plunging region is $\log\xi = 3$ in simulations D1 and D2 and $\log\xi = 4.7$ in the other simulations. The choice of the input parameters in these 8~simulations will be discussed in the next section.

The simulated observations are fitted with the model

\vspace{0.2cm}

{\sc tbabs$\times$relxill\_nk} , 

\vspace{0.2cm} 

\noindent where {\sc relxill\_nk}~\cite{Bambi:2016sac,Abdikamalov:2019yrr} is an extension of the {\sc relxill} package~\cite{Garcia:2013oma,Garcia:2013lxa} to non-Kerr spacetimes. {\sc relxill\_nk} employs a parametric black hole spacetime, namely the Kerr metric deformed by a number of deformation parameters~\cite{Vigeland:2009pr, Johannsen:2013rqa}. When all the deformation parameter vanish, we exactly recover the Kerr metric. With the spirit of a null-experiment, we can fit the data of a source to infer the values of these deformation parameter and check {\it a posteriori} whether the deformation parameters are indeed consistent with zero, as required in GR. In the present paper, we employ the Johannsen metric~\cite{Johannsen:2015pca} with the deformation parameter $\alpha_{13}$. While $\alpha_{13}$ is an {\it ad hoc} deformation parameter of the Kerr metric, an observational constraint on $\alpha_{13}$ could be translated into constraints on the coupling parameters of modified theories~\cite{Cardenas-Avendano:2019pec, Cardenas-Avendano:2019zxd}. We choose the deformation parameter $\alpha_{13}$ as it has the strongest impact, among all the deformation parameters in the Johannsen metric, on the reflection spectrum~\cite{Bambi:2016sac}.

\begin{figure*}[t]
\begin{center}
\includegraphics[type=pdf,ext=.pdf,read=.pdf,width=8.9cm]{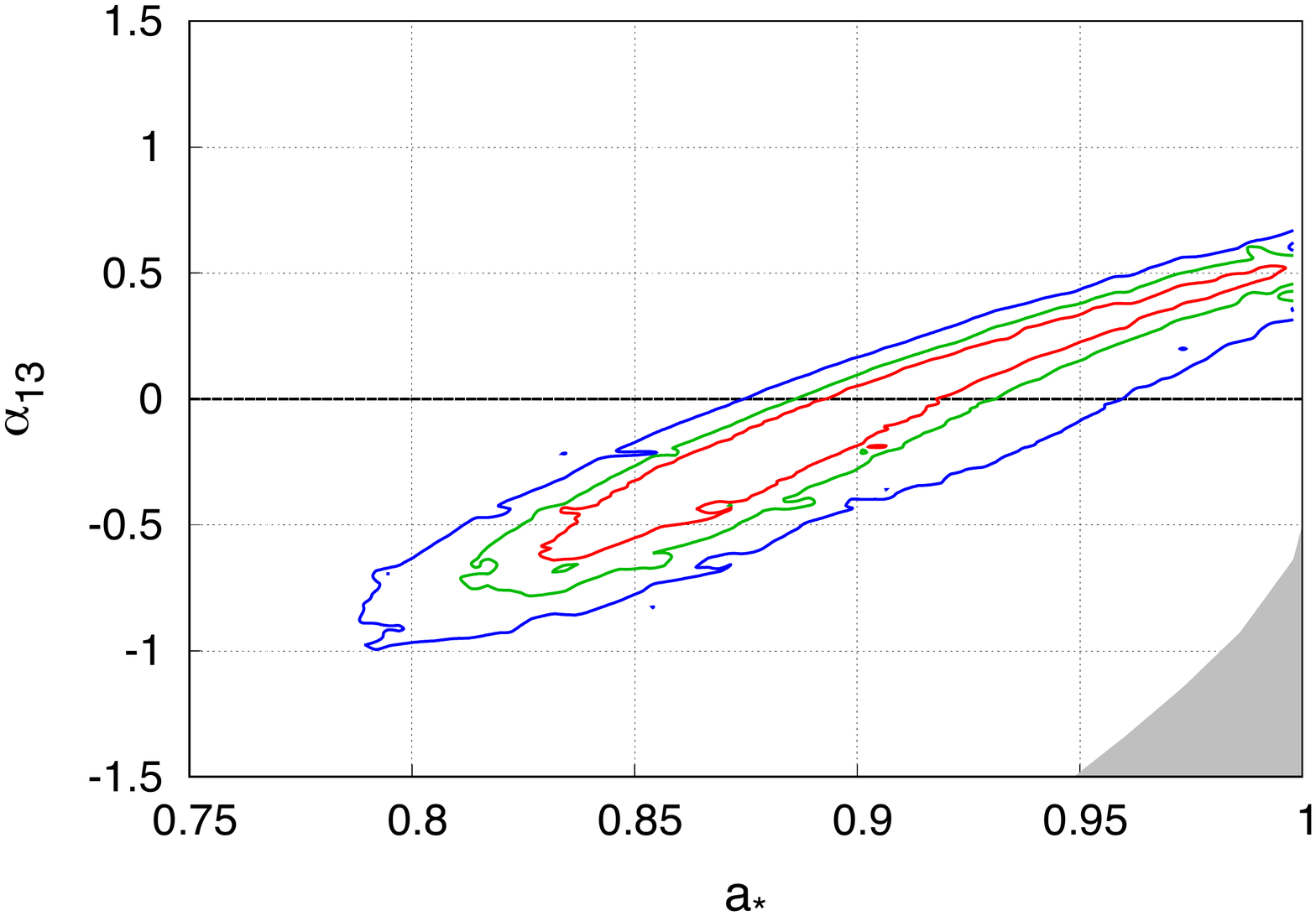}
\includegraphics[type=pdf,ext=.pdf,read=.pdf,width=8.9cm]{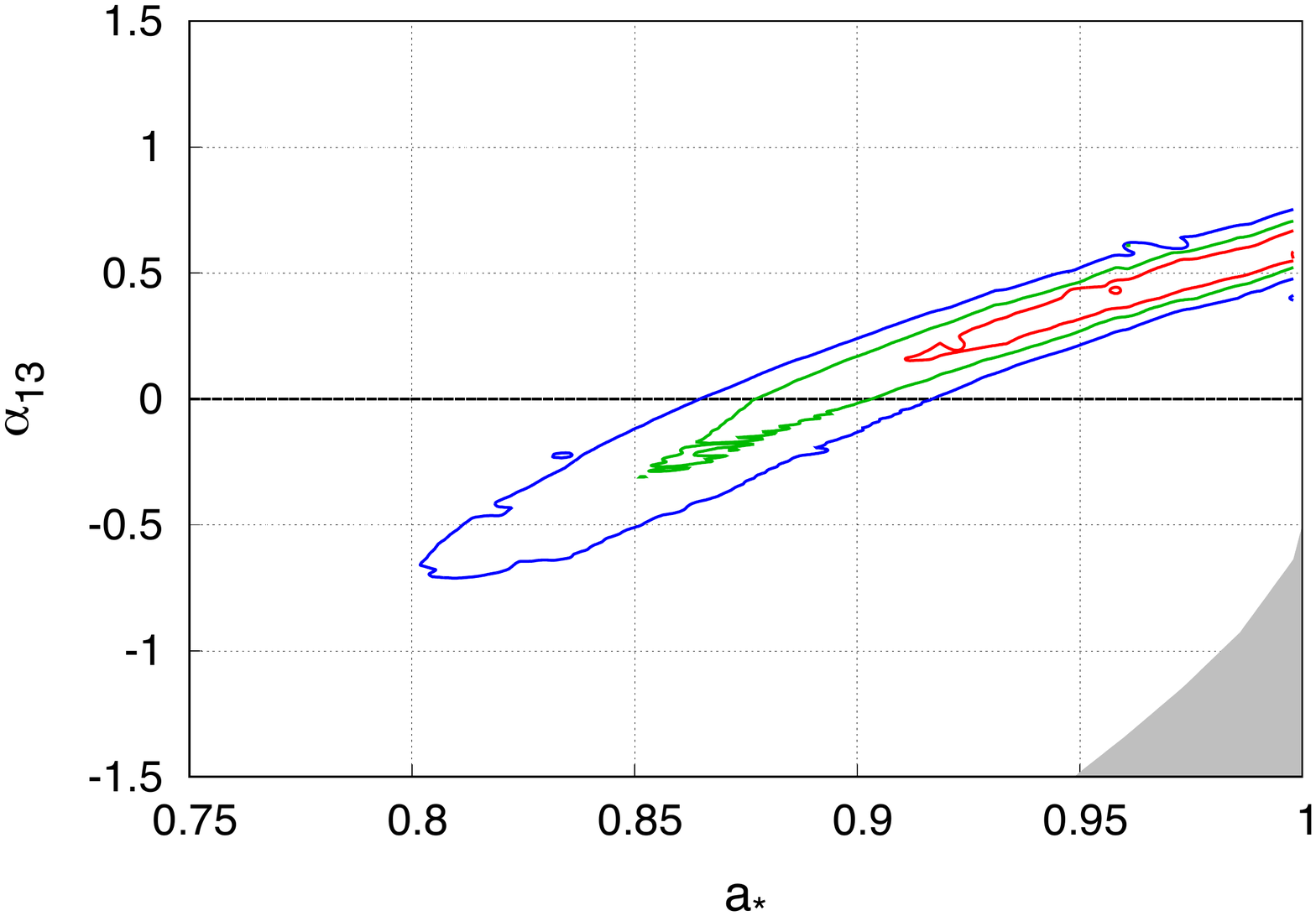}
\end{center}
\vspace{-1.3cm}
\caption{Constraints on the spin parameter $a_*$ and the deformation parameter $\alpha_{13}$ by fitting simulations A1 (left panel) and A2 (right panel) with the model {\sc tbabs$\times$relxill\_nk} that does not include the radiation from the plunging region. The input parameters were $a_* = 0.9$ and $\alpha_{13} = 0$. The input inclination angle was $i=45^{\circ}$. The red, green, and blue curves mark, respectively, the 68\%, 90\%, and 99\% confidence level bounds for two relevant parameters. The gray area in the bottom right corner is not considered in our analysis because the spacetime has pathological properties there. Note that these constraints are obtained marginalizing over all other free parameters of the fit. \label{f-45}}
\vspace{0.0cm}
\begin{center}
\includegraphics[type=pdf,ext=.pdf,read=.pdf,width=8.9cm]{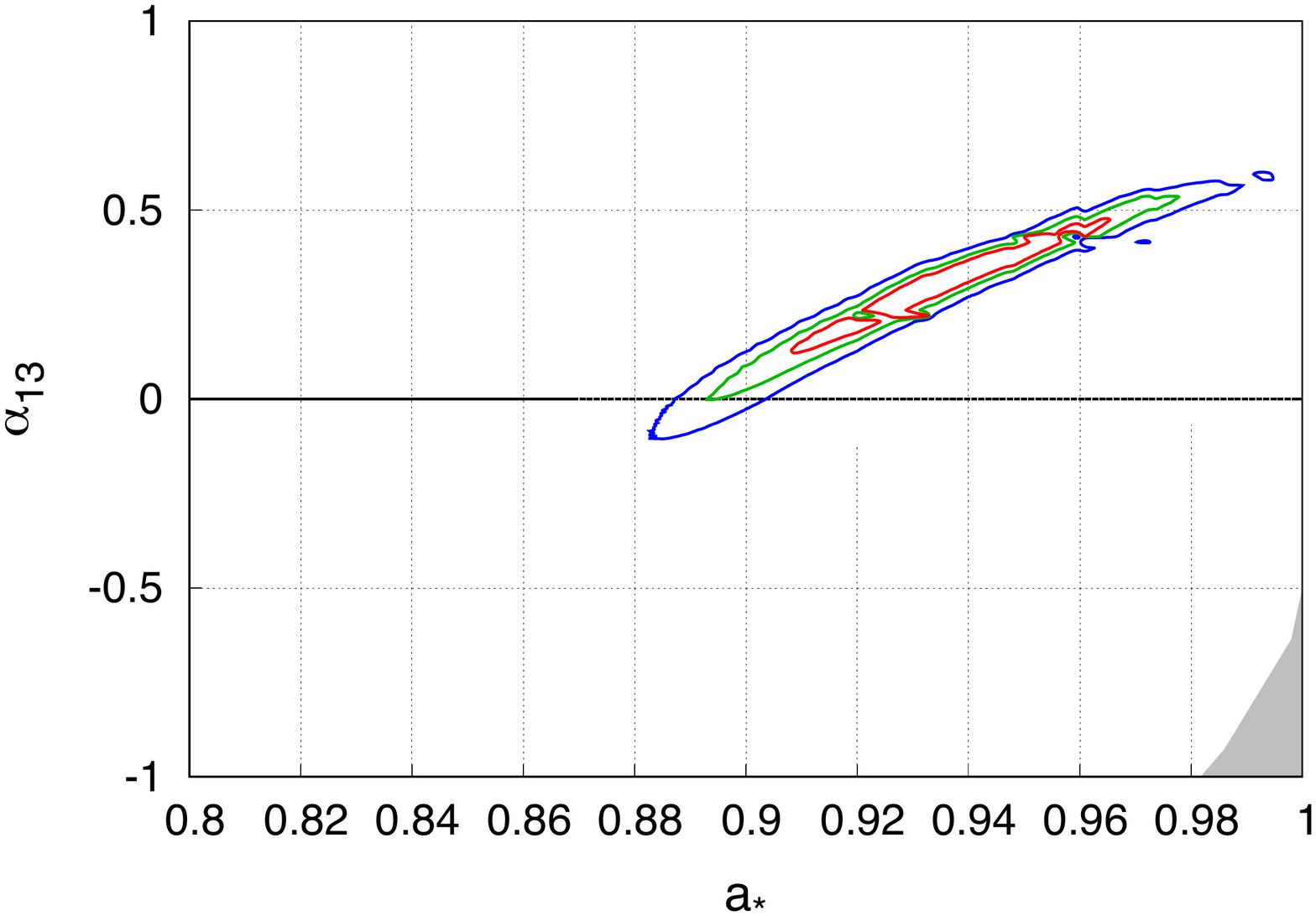}
\includegraphics[type=pdf,ext=.pdf,read=.pdf,width=8.9cm]{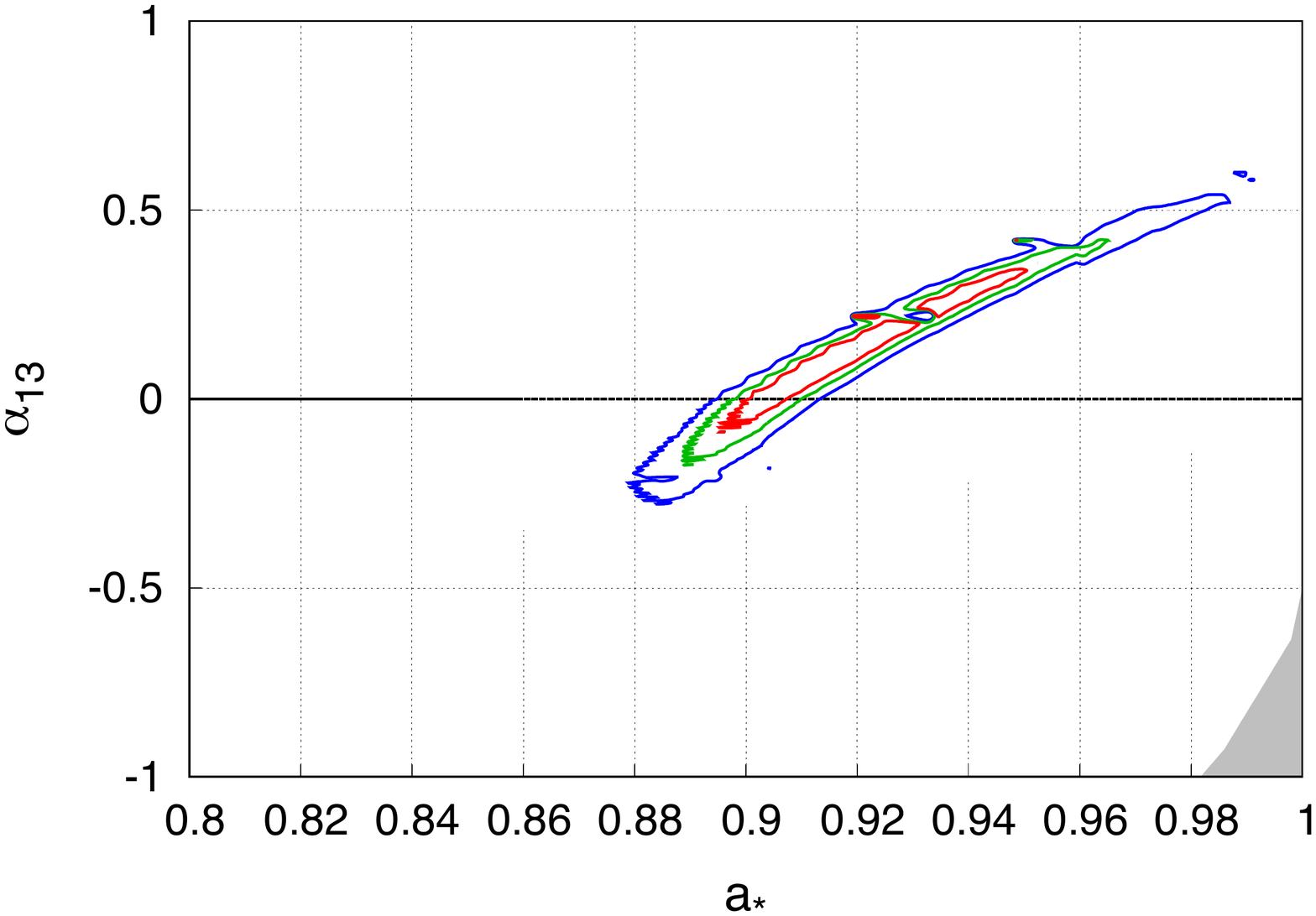}
\end{center}
\vspace{-1.3cm}
\caption{Constraints on the spin parameter $a_*$ and the deformation parameter $\alpha_{13}$ by fitting simulations B1 (left panel) and B2 (right panel) with the model {\sc tbabs$\times$relxill\_nk} that does not include the radiation from the plunging region. The input parameters were $a_* = 0.9$ and $\alpha_{13} = 0$. The input inclination angle was $i=75^{\circ}$. The red, green, and blue curves mark, respectively, the 68\%, 90\%, and 99\% confidence level bounds for two relevant parameters. The gray area in the bottom right corner is not considered in our analysis because the spacetime has pathological properties there. Note that these constraints are obtained marginalizing over all other free parameters of the fit. \label{f-75}}
\end{figure*}

\begin{figure*}[t]
\begin{center}
\includegraphics[type=pdf,ext=.pdf,read=.pdf,width=8.9cm]{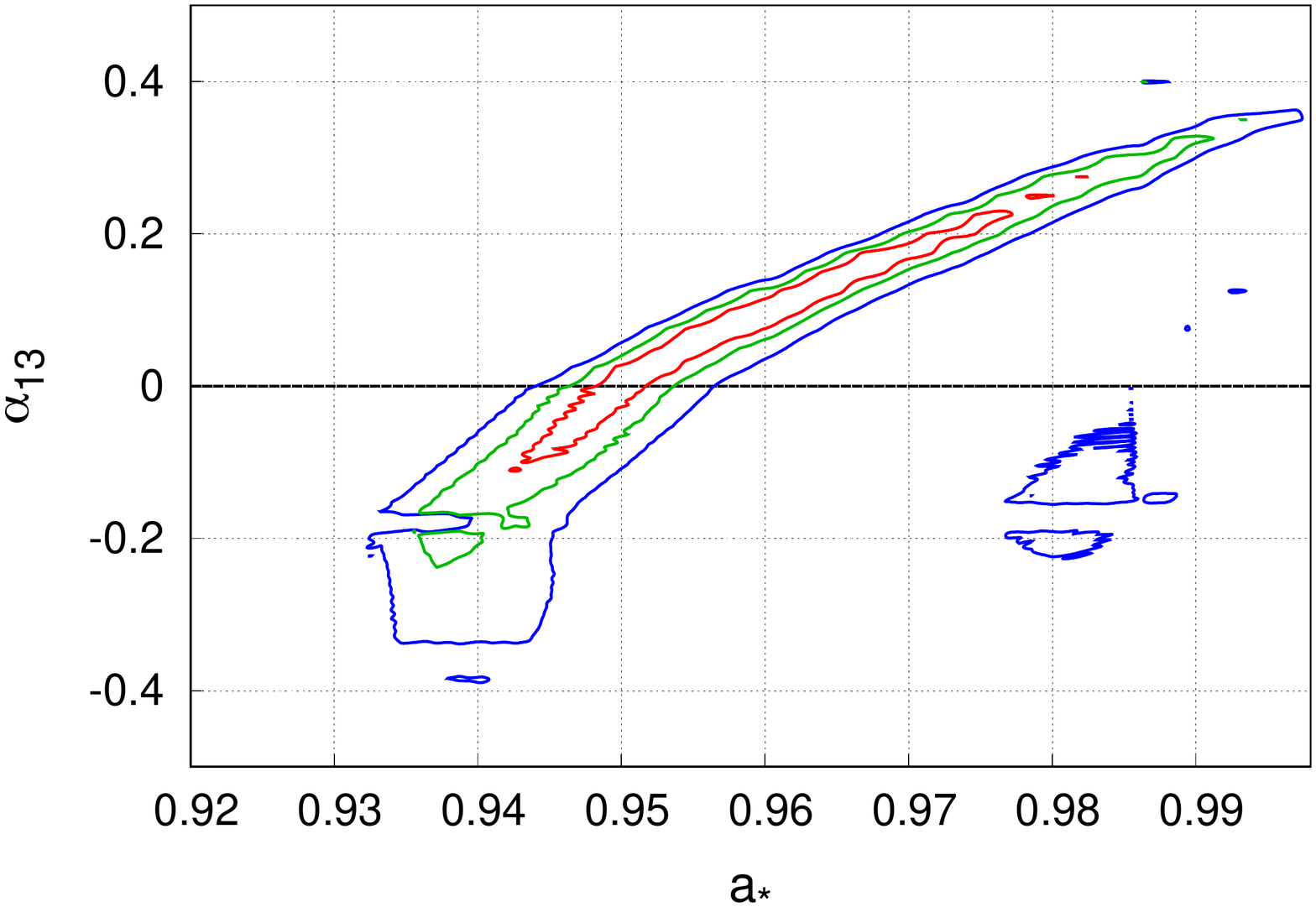}
\includegraphics[type=pdf,ext=.pdf,read=.pdf,width=8.9cm]{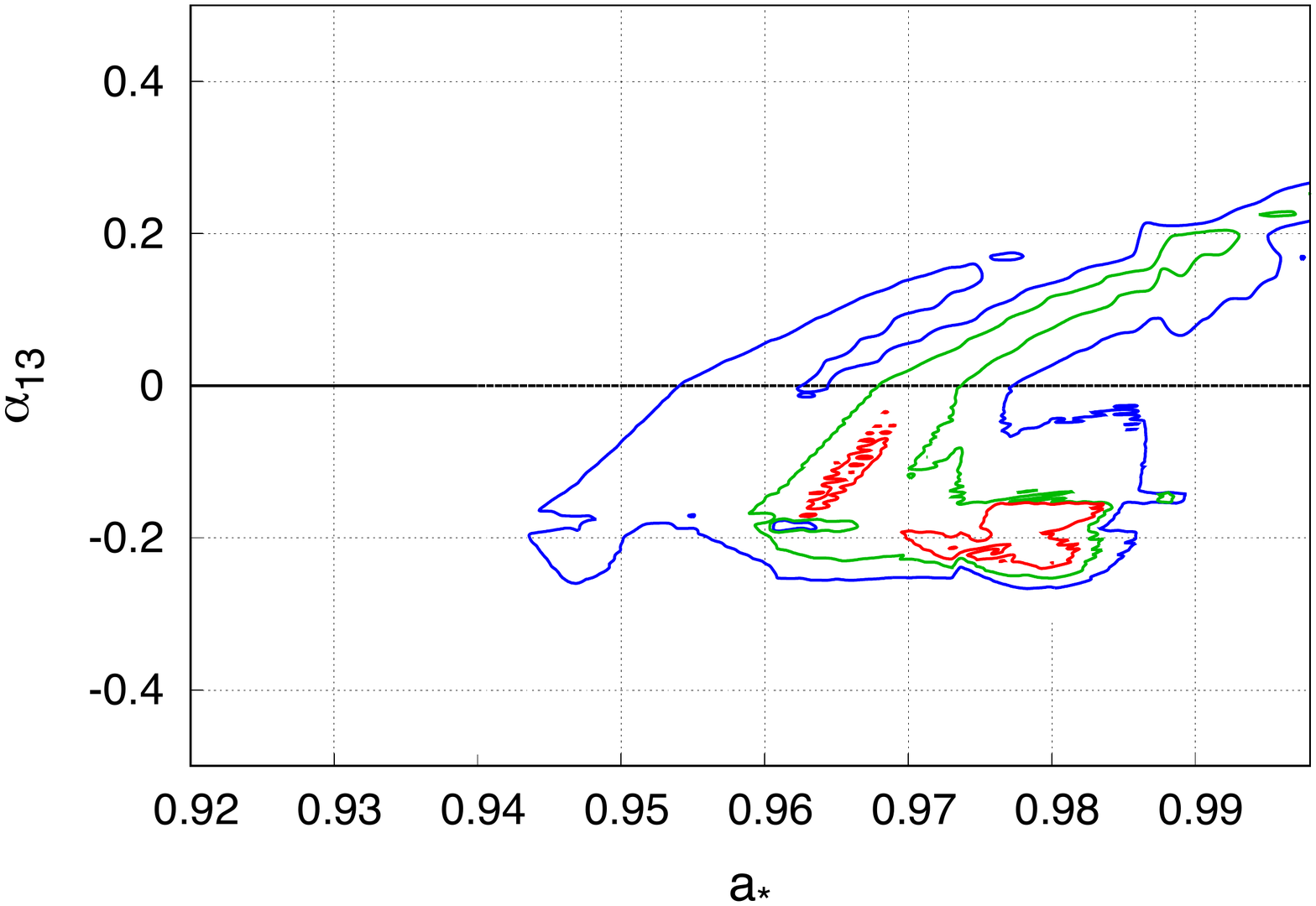}
\end{center}
\vspace{-1.3cm}
\caption{Constraints on the spin parameter $a_*$ and the deformation parameter $\alpha_{13}$ by fitting simulations C1 (left panel) and C2 (right panel) with the model {\sc tbabs$\times$relxill\_nk} that does not include the radiation from the plunging region. The input parameters were $a_* = 0.99$ and $\alpha_{13} = 0$. The input ionization parameter in the plunging region was $\log\xi=4.7$. The input inclination angle was $i=75^{\circ}$. The red, green, and blue curves mark, respectively, the 68\%, 90\%, and 99\% confidence level bounds for two relevant parameters. Note that these constraints are obtained marginalizing over all other free parameters of the fit. \label{f-99a}}
\vspace{-0.4cm}
\begin{center}
\includegraphics[type=pdf,ext=.pdf,read=.pdf,width=8.9cm]{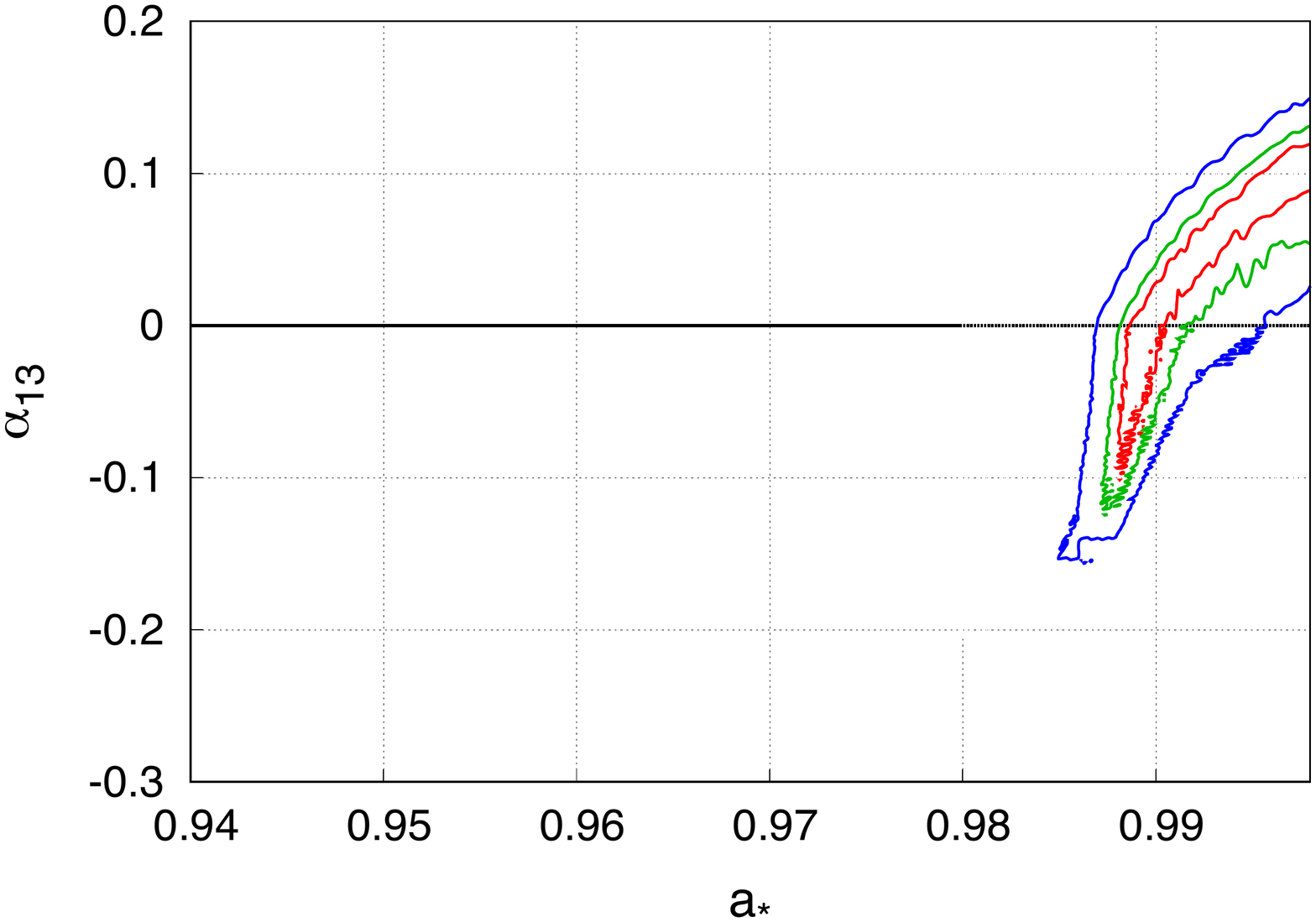}
\includegraphics[type=pdf,ext=.pdf,read=.pdf,width=8.9cm]{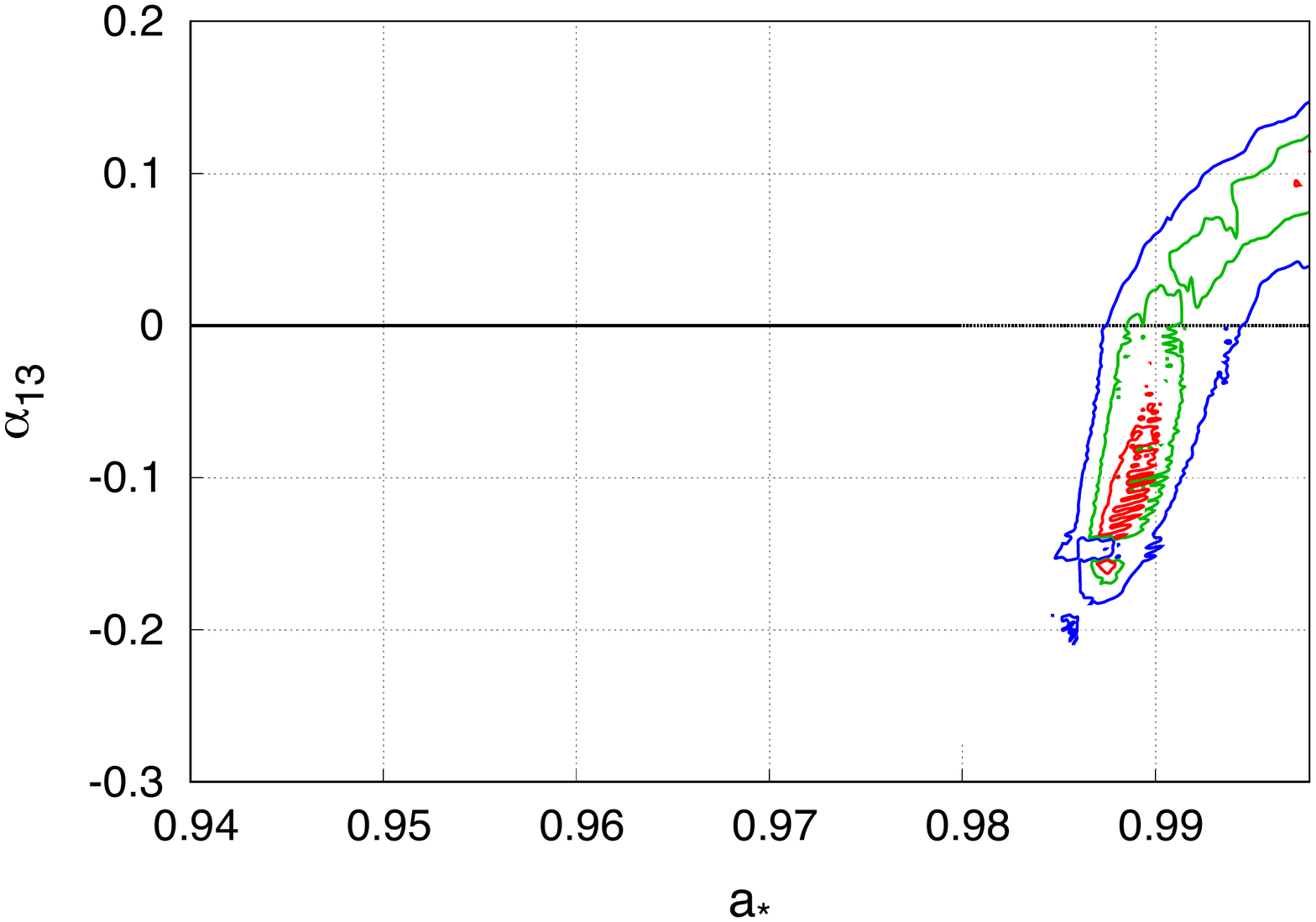}
\end{center}
\vspace{-1.3cm}
\caption{Constraints on the spin parameter $a_*$ and the deformation parameter $\alpha_{13}$ by fitting simulations D1 (left panel) and D2 (right panel) with the model {\sc tbabs$\times$relxill\_nk} that does not include the radiation from the plunging region. The input parameters were $a_* = 0.99$ and $\alpha_{13} = 0$. The input ionization parameter in the plunging region was $\log\xi=3$. The input inclination angle was $i=75^{\circ}$. The red, green, and blue curves mark, respectively, the 68\%, 90\%, and 99\% confidence level bounds for two relevant parameters. Note that these constraints are obtained marginalizing over all other free parameters of the fit. \label{f-99b}}
\end{figure*}

The results of our fits for simulations A1, A2, B1, and B2 are reported in Tab.~\ref{t-fit45} (for A1 and A2) and Tab.~\ref{t-fit75} (for B1 and B2). Note that for every simulation we consider two models: we assume GR and we set $\alpha_{13} = 0$, and we do not assume GR and $\alpha_{13}$ is a free parameter in the fit. The constraints on the plane black hole spin parameter vs deformation parameter are shown in Fig.~\ref{f-45} (for A1 and A2) and Fig.~\ref{f-75} (for B1 and B2). The best-fit values inferred from simulations C1, C2, D1, and D2 are shown in Tab.~\ref{t-fit99}, where this time we only report the results of the non-GR fit with $\alpha_{13}$ free. The constraints on $a_*$ and $\alpha_{13}$ are shown in Fig.~\ref{f-99a} (for C1 and C2) and in Fig.~\ref{f-99b} (for D1 and D2). In all figures, the red, green, and blue curves are for the 68\%, 90\%, and 99\% confidence level limits for two relevant parameters. The gray area in the bottom right corner of Fig.~\ref{f-45} and Fig.~\ref{f-75} is ignored in our analysis because it is a region of the parameter space were pathological behavior appears~\cite{Bambi:2016sac}. The features of some confidence level curves are non-physical and related to the difficulties of the algorithm of XSPEC to map properly the parameter space to find the minimum of $\chi^2$ marginalized over all the other free parameters of the fit~\cite{Cardenas-Avendano:2019pec}.  

\begin{table*}
\centering
{\renewcommand{\arraystretch}{1.3}
\begin{tabular}{cc|cc|cc}
\hline 
 & & \multicolumn{2}{c}{A1} & \multicolumn{2}{c}{A2} \\
 & \hspace{0.2cm} Input \hspace{0.2cm} & GR & non-GR & GR & non-GR \\
\hline 
{\sc tbabs} &&&&& \\
$N_{\rm H}/10^{20}$~cm$^{-2}$ & $6.74$ & $6.74^{*}$ & $6.74^{*}$ & $6.74^{*}$ & $6.74^{*}$ \\
\hline 
{\sc relxill\_nk} &&&&& \\
$q$ & $3$ & $2.93_{-0.04}^{+0.05}$ & $2.92_{-0.04}^{+0.05}$ & $2.95_{-0.03}^{+0.03}$ & $2.972_{-0.020}^{+0.040}$ \\
$i$ {[}deg{]}  & $45$ & $44.35_{-0.07}^{+0.07}$ & $44.40_{-0.06}^{+0.06}$ & $44.32_{-0.07}^{+0.08}$ & $44.22_{-0.05}^{+0.04}$\\
$a_*$ & $0.9$ & $0.907_{-0.018}^{+0.017}$ & $0.88_{-0.05}^{+0.10}$ & $0.890_{-0.012}^{+0.011}$ & $0.96_{-0.06}^{\rm + (P)}$\\
$\log\xi$ & $3$ & $3.016_{-0.003}^{+0.003}$ & $3.016_{-0.003}^{+0.003}$ & $3.0139_{-0.0019}^{+0.0023}$ & $3.0144_{-0.0015}^{+0.0014}$\\
$A_{\rm Fe}$ & $1$ & $1.136_{-0.020}^{+0.020}$ & $1.136_{-0.018}^{+0.014}$ & $1.155_{-0.014}^{+0.013}$ & $1.157_{-0.014}^{+0.015}$\\
$\Gamma$ & $2$ & $2.0122_{-0.0017}^{+0.0015}$ & $2.0120_{-0.0015}^{+0.0013}$ & $2.0109_{-0.0009}^{+0.0011}$ & $2.0112_{-0.0006}^{+0.0005}$\\
$E_{\rm cut}$ [keV] & 150 & 150$^*$ & 150$^*$ & 150$^*$ & 150$^*$ \\
\hline 
$\alpha_{13}$ & $0$ & $0^{*}$ & $-0.2_{-0.4}^{+0.7}$ & $0^{*}$ & $0.4_{-0.3}^{+0.3}$\tabularnewline
\hline 
$\chi^{2}/\nu$ &  & \hspace{0.2cm} $20042.69/20412$ \hspace{0.2cm} & \hspace{0.2cm} $20042.18/20411$ \hspace{0.2cm} & \hspace{0.2cm} $20186.14/20423$ \hspace{0.2cm} & \hspace{0.2cm} $20184.37/20422$\hspace{0.2cm} \\
 &  & $=0.9819074$ & $=0.9819303$ & $=0.9884024$ & $=0.9883641$\tabularnewline
\hline 
\end{tabular}
}
\caption{Input parameters and best-fit values for simulations A1 and A2. For both simulations, we fitted the data with the GR model ($\alpha_{13} = 0$) and non-GR model ($\alpha_{13}$ free in the fit). The reported uncertainties correspond to 90\% confidence level for one relevant parameter. $^\star$ indicates that the parameter is frozen in the fit. (P) indicates that the 90\% confidence level uncertainty reaches the boundary of the parameter space (for the spin parameter, the upper boundary is at $a_* = 0.998$).} \label{t-fit45}
\vspace{0.8cm}
\centering
{\renewcommand{\arraystretch}{1.3}
\begin{tabular}{cc|cc|cc}
\hline 
 & & \multicolumn{2}{c}{B1} & \multicolumn{2}{c}{B2} \\
 & \hspace{0.2cm} Input \hspace{0.2cm} & GR & non-GR & GR & non-GR \\
\hline 
{\sc tbabs} &&&&& \\
$N_{\rm H}/10^{20}$~cm$^{-2}$ & $6.74$ & $6.74^{*}$ & $6.74^{*}$ & $6.74^{*}$ & $6.74^{*}$ \\
\hline 
{\sc relxill\_nk} &&&&& \\
$q$ & $3$ & $3.122_{-0.031}^{+0.023}$ & $3.177_{-0.021}^{+0.029}$ & $3.049_{-0.016}^{+0.022}$ & $3.072_{-0.019}^{+0.022}$\tabularnewline
$i$ {[}deg{]}  & $75$ & $74.56_{-0.06}^{+0.09}$ & $73.9_{-0.4}^{+0.4}$ & $74.70_{-0.05}^{+0.10}$ & $74.4_{-0.5}^{+0.5}$\\
$a_*$ & $0.9$ & $0.894_{-0.005}^{+0.008}$ & $0.94_{-0.03}^{+0.03}$ & $0.903_{-0.004}^{+0.005}$ & $0.92_{-0.03}^{+0.03}$\\
$\log\xi$ & $3.0$ & $3.029_{-0.006}^{+0.003}$ & $3.0306_{-0.0024}^{+0.0051}$ & $3.0216_{-0.0022}^{+0.0009}$ & $3.0220_{-0.0048}^{+0.0021}$\\
$A_{\rm Fe}$ & $1$ & $1.103_{-0.023}^{+0.013}$ & $1.114_{-0.014}^{+0.024}$ & $1.090_{-0.012}^{+0.012}$ & $1.094_{-0.012}^{+0.012}$\\
$\Gamma$ & $2$ & $2.0207_{-0.0018}^{+0.0014}$ & $2.0207_{-0.0016}^{+0.0016}$ & $2.0158_{-0.0004}^{+0.0007}$ & $2.0157_{-0.0017}^{+0.0011}$\\
$E_{\rm cut}$ [keV] & 150 & 150$^*$ & 150$^*$ & 150$^*$ & 150$^*$ \\
\hline 
$\alpha_{13}$ & $0$ & $0^{*}$ & $0.3_{-0.2}^{+0.2}$ & $0^{*}$ & $0.1_{-0.2}^{+0.3}$\tabularnewline
\hline 
$\chi^{2}/\nu$ &  & \hspace{0.2cm} $20285.30/20333$ \hspace{0.2cm} & \hspace{0.2cm} $20280.79/20332$ \hspace{0.2cm} & \hspace{0.2cm} $19899.69/20341$ \hspace{0.2cm} & \hspace{0.2cm} $19899.05/20340$\\
 &  & $=0.9976539$ & $=0.9974816$ & $=0.9783044$ & $=0.9783209$\\
\hline 
\end{tabular}}
\caption{Input parameters and best-fit values for simulations B1 and B2. For both simulations, we fitted the data with the GR model ($\alpha_{13} = 0$) and non-GR model ($\alpha_{13}$ free in the fit). The reported uncertainties correspond to 90\% confidence level for one relevant parameter. $^\star$ indicates that the parameter is frozen in the fit.} \label{t-fit75}
\end{table*}

\begin{table*}
\centering
{\renewcommand{\arraystretch}{1.3}
\begin{tabular}{cc|c|c|c|c}
\hline 
 & & C1 & C2 & D1 & D2 \\
 & \hspace{0.2cm} Input \hspace{0.2cm} &&&& \\
\hline 
{\sc tbabs} &&&&& \\
$N_{\rm H}/10^{20}$~cm$^{-2}$ & $6.74$ & $6.74^{*}$ & $6.74^{*}$ & $6.74^{*}$ & $6.74^{*}$ \\
\hline 
{\sc relxill\_nk} &&&&& \\
$q$ & $3$ & $2.755_{-0.016}^{+0.024}$ & $3.03_{-0.09}^{+0.04}$ & $3.20_{-0.06}^{+0.08}$ & $3.22_{-0.11}^{+0.18}$ \\
$i$ {[}deg{]}  & $75$ & $74.21_{-0.12}^{+0.49}$ & $75.53_{-0.07}^{+0.08}$ & $74.89_{-0.04}^{+0.03}$ & $75.31_{-0.40}^{+0.06}$\\
$a_*$ & $0.99$ & $0.96_{-0.02}^{+0.02}$ & $0.979_{-0.018}^{+0.004}$ & $0.992_{-0.005}^{\rm + (P)}$ & $0.9879_{-0.0003}^{\rm + (P)}$ \\
$\log\xi$ & $3$ & $3.022_{-0.003}^{+0.004}$ & $3.0122_{-0.0022}^{+0.0018}$ & $3.0179_{-0.0023}^{+0.0031}$ & $3.0124_{-0.0022}^{+0.0016}$ \\
$A_{\rm Fe}$ & $1$ & $1.094_{-0.014}^{+0.009}$ & $1.089_{-0.010}^{+0.009}$ & $1.079_{-0.010}^{+0.010}$ & $1.078_{-0.010}^{+0.009}$ \\
$\Gamma$ & $2$ & $2.0160_{-0.0018}^{+0.0018}$ & $2.0131_{-0.0017}^{+0.0014}$ & $2.0132_{-0.0008}^{+0.0016}$ & $2.0114_{-0.0014}^{+0.0015}$ \\
$E_{\rm cut}$ [keV] & 150 & 150$^*$ & 150$^*$ & 150$^*$ & 150$^*$ \\
\hline 
$\alpha_{13}$ & $0$ & $0.1_{-0.2}^{+0.2}$ & $-0.18_{-0.15}^{+0.2}$ & $0.05_{-0.15}^{+0.07}$ & $-0.1_{-0.1}^{+0.2}$\\
\hline 
$\chi^{2}/\nu$ &  & \hspace{0.2cm} $20099.05/20344$ \hspace{0.2cm} & \hspace{0.2cm} $20184.77/20350$ \hspace{0.2cm} & \hspace{0.2cm} $20146.53/20346$ \hspace{0.2cm} & \hspace{0.2cm} $19907.64/20357$ \hspace{0.2cm} \\
 &  & $=0.9879596$ & $=0.9918807$ & $=0.9901961$ & $=0.9779258$\\
\hline 
\end{tabular}
}
\caption{Input parameters and best-fit values for simulations C1, C2, D1, and D2. Here we only show the non-GR model with $\alpha_{13}$ free in the fit. The reported uncertainties correspond to 90\% confidence level for one relevant parameter. $^\star$ indicates that the parameter is frozen in the fit. (P) indicates that the 90\% confidence level uncertainty reaches the boundary of the parameter space (for the spin parameter, the upper boundary is at $a_* = 0.998$).} \label{t-fit99}
\end{table*}


\section{Discussion and conclusions}
\label{Conclusions}

We have calculated single line shapes and full reflection spectra with and without the contribution from the plunging region, and we have found that the impact of the radiation from the plunging gas is more important for low and negative values of the black hole spin parameters and high values of the viewing angles. We can thus expect that systematic uncertainties and modeling bias are larger for sources with low values of $a_*$ observed with high values of $i$. However, X-ray reflection spectroscopy is thought to provide reliable measurements for high values of the black hole spin parameter, say $a_* > 0.9$, while it is often difficult to break the parameter degeneracy in the case of sources with low or moderate values of $a_*$~\cite{Dauser:2013xv, Kammoun:2018ltv,Cardenas-Avendano:2019pec}. This point is even more important if we want to test the Kerr metric of the source, where more stringent constraints require that the inner edge of the accretion disk is as close as possible to the black hole event horizon. For this reason, in our first 4~simulations (A1, A2, B1, and B2) we assumed $a_* = 0.9$: for lower values of the black hole spin it is more and more difficult to recover the correct properties of the system independent of the modeling uncertainties and they are not our target sources for testing the Kerr black hole hypothesis. Generally speaking, simulations B1 and B2 should represent the worse situations with $a_* = 0.9$ and $i = 75^\circ$.

From our previous work on tests of the Kerr metric with black hole binaries and AGNs, the most stringent constraints on the Johannsen deformation parameter $\alpha_{13}$ have been obtained from the analysis of a 117~ks \textit{Suzaku} observation in 2007 of the black hole binary GRS~1915+105. Our spectral analysis with {\sc relxill\_nk} gave (90\% of confidence level for two relevant parameters)~\cite{Zhang:2019ldz,Abdikamalov:2020oci}
\be\label{eq-grs}
a_* > 0.988 \, , \quad -0.25 < \alpha_{13} < 0.08 \, .
\ee
and an inclination angle of the disk $i \sim 75^\circ$. We have thus decided to study better the impact of the radiation from the plunging region on the reflection spectrum of a source and we have designed simulations C1, C2, D1, and D2. All these simulations have an input spin parameter $a_* = 0.99$ and viewing angle $i = 75^\circ$. Considering that the ionization parameter $\xi$ does not have discontinuities at the ISCO~\cite{Wilkins:2020pgu}, we have decided to explore the two extreme cases for the ionization in the plunging region, namely $\log\xi = 3$ as in the disk and $\log\xi = 4.7$ which is the maximum value allowed in {\sc xillver}.

All the simulations have been done with the XIFU instrument onboard of \textsl{Athena}, which is expected to be launched after 2032. The XIFU instrument has an excellent energy resolution (2.5~eV near the iron line, to be compared with current instruments on board of \textsl{XMM-Newton} with an energy resolution $\sim 150$~eV near the iron line) and a larger effective area than current instruments. We are thus considering the situation of an optimistic observation, not possible today: if we find that the impact of the radiation from the plunging region is weak for a similar observation, we can argue that its impact should be even weaker for the available X-ray data.

In general, the fits of our simulations show that all the model parameters can be recovered even if we do not employ a model that takes the radiation from the plunging region into account, and the choice of the intensity profile in the plunging region, either $\epsilon^{\,(1)}$ or $\epsilon^{\,(2)}$, does not seem to play any significant role, which is a good news for us because we do not know it. There are some minor biases. For example, the iron abundance $A_{\rm Fe}$, the ionization parameter of the disk $\log\xi$, and the photon index $\Gamma$ are always slightly overestimates, no matter if we assume GR or not. However, while we do not recover the correct input parameters with the 90\% confidence level for $A_{\rm Fe}$, $\log\xi$, and $\Gamma$, the discrepancies with the input values are small.

When we test the Kerr black hole hypothesis, the two important parameters are the black hole spin $a_*$ and the deformation parameter $\alpha_{13}$, and for this reason we have reported their constraints in Figs.~\ref{f-45}-\ref{f-99b}. The typical banana shape of the confidence level contours is related to the well-known fact that these two parameters are usually correlated. In simulations A1, A2, B1, and B2 with $a_* = 0.9$, we recover the input parameter point $(a_* ; \alpha_{13})=(0.9 ; 0)$. In the other simulations with input $a_* = 0.99$, we may not recover the correct input spin parameter (this is the case of C1), but still we recover $\alpha_{13} = 0$. Our conclusion is that tests of the Kerr black hole hypothesis, like that in Eq.~(\ref{eq-grs}) using the black hole binary GRS~1915+105, are not appreciably affected by the reflection radiation produced in the plunging region.

We can compare our results with those already present in literature, even if our study is mainly focused on the evaluation of the impact of the radiation from the plunging region on the possibility of testing the Kerr metric while the other studies in literature always assume the Kerr metric. In Ref.~\cite{Reynolds:2007rx}, the authors present the results of numerical simulations of geometrically thin accretion disk in a pseudo-Newtonian potential. They find that the radiation from the plunging region affects the estimate of the black hole spin, with larger systematic errors for slow-rotating black holes and and smaller and smaller errors as the black hole spin parameter $a_*$ approaches to 1. This is qualitatively what we see in our Fig.~\ref{f-full}, which can be easily interpreted with the fact that the plunging region is larger when the ISCO radius is larger. A more recent study of the impact of the radiation from the plunging region is reported in Ref.~\cite{Wilkins:2020pgu}, where the authors study how X-ray reverberation mapping can provide information regarding the presence of the ISCO. They show that the contribution from the plunging region has a minimal effect on the time-averaged X-ray spectrum and the overall lag-energy spectrum, still in agreement with our results. They show that the plunging region can be distinguished from the disk emission due to the rapid increase of the ionization in the plunging region, as well as the intrinsic energy shifts that appear there.

Lastly, we note that the size of the plunging region, as well as the motion of the gas in the plunging region and in the inner part of the accretion disk, may be significantly altered by the presence of magnetic fields, which are completely ignored in the present study and will be investigated in future work. In the case of geometrically thin disks (which is the case studied here), the impact of magnetic fields may be negligible~\cite{Penna:2010hu,Penna:2011rw}, but it is not in the presence of a highly magnetized coronal region around the black hole~\cite{Noble:2010mm}. A preliminary study on the iron line shape from magnetized disks is reported in Ref.~\cite{Frolov:2014zia}.

\acknowledgments

We gratefully acknowledge Dimitry Ayzenberg, Javier Garc\'ia, Jiachen Jiang, and Andrea Lopera for useful discussions and comments. A.C.-A. acknowledges funding from the Fundaci\'on Universitaria Konrad Lorenz (Project 5INV1). The work of M.Z. and C.B. was supported by the Innovation Program of the Shanghai Municipal Education Commission, Grant No.~2019-01-07-00-07-E00035, and the National Natural Science Foundation of China (NSFC), Grant No.~11973019. A.C.-A. also wishes to thank the Department of Physics at Fudan University, where part of this work was performed, for their hospitality. Computational efforts were performed on the Scientific Computing Laboratory, operated and supported by Fundaci\'on Universitaria Konrad Lorenz’s Engineering and Mathematics Department.


\bibliography{References}

\end{document}